\documentclass[%
showpacs,preprintnumbers,
nobibnotes,
 amsmath,amssymb,
 aps,
 pre,
 longbibliography,
 lengthcheck,%
]{revtex4-1}

\usepackage{graphicx}% Include figure files
\usepackage{dcolumn}% Align table columns on decimal point
\usepackage{bm}% bold math
\usepackage{hyperref}% add hypertext capabilities

\usepackage{lineno}

\begin{document}

\title{Scarce defects induce anomalous deterministic diffusion}

\author{M. Hidalgo-Soria} 
\affiliation{Centro de Investigaci\'on en Ciencias-IICBA, Universidad Aut\'onoma del Estado de Morelos. Avenida Universidad 1001, Colonia Chamilpa, 62209, Cuernavaca Morelos, Mexico.} 

\author{R. Salgado-Garc\'{\i}a} 
\email{Corresponding author: raulsg@uaem.mx}
\affiliation{Centro de Investigaci\'on en Ciencias-IICBA, Universidad Aut\'onoma del Estado de Morelos. Avenida Universidad 1001, Colonia Chamilpa, 62209, Cuernavaca Morelos, Mexico.} 
\date{\today}

\begin{abstract}

We introduce a simple model of deterministic particles in weakly disordered media which exhibits a transition from normal to anomalous diffusion. The model consists of a set of non-interacting overdamped particles moving on a disordered potential. The disordered potential can be thought as a substrate having some ``defects'' scattered along a one-dimensional line. The distance between two contiguous defects is assumed to have a heavy-tailed distribution with a given exponent $\alpha$, which means that the defects along the substrate are scarce if $\alpha$ is small. We prove that this system exhibits a transition from normal to anomalous diffusion when the distribution exponent $\alpha$  decreases, i.e., when the defects become scarcer.  Thus we identify three distinct scenarios: a normal diffusive phase for $\alpha>2$, a superdiffusive phase for $1/2<\alpha\leq 2$, and a subdiffusive phase for $\alpha \leq 1/2$. 
We also prove that the particle current is finite for all the values of $\alpha$, which means that the transport is normal independently of the diffusion regime (normal, subdiffusive, or superdiffusive). 
We give analytical expressions for the effective diffusion coefficient for the normal diffusive phase and analytical expressions for the diffusion exponent in the case of anomalous diffusion. We test all these predictions by means of numerical simulations.

\end{abstract}


%% Creating title
\maketitle

%%%%%%%%%%%%%%%%%%%%%%%%%%%%%%%%%%%%%%%%%%%%%%%%%%%%%%%%%%%%%%%%%%%%%%%%%%%%%%
%%%%%%%%%%%%%%%%%%%%%%%%%%%%%%%%%%%%%%%%%%%%%%%%%%%%%%%%%%%%%%%%%%%%%%%%%%%%%%
%
\section{Introduction}
%
%%%%%%%%%%%%%%%%%%%%%%%%%%%%%%%%%%%%%%%%%%%%%%%%%%%%%%%%%%%%%%%%%%%%%%%%%%%%%%
%%%%%%%%%%%%%%%%%%%%%%%%%%%%%%%%%%%%%%%%%%%%%%%%%%%%%%%%%%%%%%%%%%%%%%%%%%%%%%

In recent years there have been an increasing interest in a singular phenomenon which is enhancement of diffusion by weak disorder~\cite{reimann2008weak,khoury2011weak,lindenberg2012weak,simon2013transport,salgado2014effective}. This phenomenon has been shown to occur in a system consisting of an ensemble of non-interacting overdamped particles moving on a weakly disordered periodic potential with a constant driving force in presence of Gaussian white noise~\cite{reimann2008weak}. Thereafter, it was found that the diffusion of particles in such kind of systems can become anomalous, both, superdiffusive and subdiffusive in a wide range of the parameter space~\cite{khoury2011weak}. On the other hand, in purely deterministic systems, i.e., in systems without noise fluctuations, the anomalous diffusion has also been found~\cite{kunz2003mechanical,denisov2010biased,denisov2010anomalous,salgado2013normal}.  However, contrary to the systems with noise in which the anomalous phase is robust with respect to other parameters, in deterministic systems the anomalous phase emerge as a critical property~\cite{kunz2003mechanical,denisov2010biased,denisov2010anomalous,salgado2013normal}. This means that, for deterministic systems, one of the parameter should have a critical value (the driving constant force) in order for the system to exhibits the asymptotic anomalous behavior. These finding would suggest that the origin of the anomalous diffusion in the models presented in Ref.~\cite{khoury2011weak} could be due, besides to the long-range correlation of the disordered potential~\cite{khoury2011weak}, to the presence of noise. In this work we show that this is not necessarily the case. Indeed we introduce a simple model for deterministic diffusion which exhibits a transition from normal to anomalous diffusion as a function of a parameter. This model has significant differences with respect to previously proposed models for deterministic diffusion in disordered systems. Particularly, we found that in our model the anomalous phase does not emerge as a critical property, which means that we do not require a fine-tuning of the parameters to obtain anomalous diffusion. Moreover, we show that the anomalous behavior is robust with respect to a driving constant force. These findings thus provide a different mechanism leading to anomalous deterministic  diffusion in disordered systems. 

The paper is organized as follows. In Section~\ref{sec:model} we state the model for the disordered potential. In Section~\ref{sec:anomalous} we perform the calculations to obtain the diffusion coefficient when the normal diffusion occurs. We also obtain the diffusion exponent for the anomalous phase and we prove that the systems transits from anomalous superdiffusion to subdiffusion. In Section~\ref{sec:simulations} we test our findings with numerical simulations.  Finally in Section~\ref{sec:conclusions} we give the conclusions of our work.

%%%%%%%%%%%%%%%%%%%%%%%%%%%%%%%%%%%%%%%%%%%%%%%%%%%%%%%%%%%%%%%%%%%%%%%%%%%%%%
%%%%%%%%%%%%%%%%%%%%%%%%%%%%%%%%%%%%%%%%%%%%%%%%%%%%%%%%%%%%%%%%%%%%%%%%%%%%%%
%
\section{Model}
\label{sec:model}
%
%%%%%%%%%%%%%%%%%%%%%%%%%%%%%%%%%%%%%%%%%%%%%%%%%%%%%%%%%%%%%%%%%%%%%%%%%%%%%%
%%%%%%%%%%%%%%%%%%%%%%%%%%%%%%%%%%%%%%%%%%%%%%%%%%%%%%%%%%%%%%%%%%%%%%%%%%%%%%

Let us consider an ensemble of non-interacting overdamped particles moving on a one-dimensional substrate. The equation of motion of each particle is given by
\begin{equation}
\gamma \frac{dx}{dt} = f(x) + F,
\label{eq:motion}
\end{equation}
where $f(x)$ is minus the gradient of a potential $V(x)$ and $F$ is constant driving force. We assume that $V(x)$ is a weakly disordered potential in the sense that it consists of some ``defects'' scattered along the substrate. In order to write an analytical expression for $V(x)$ let us introduce a function defined on a finite interval that will play the role of defect. Let $\varphi : [0,L] \to \mathbb{R}$ be a real-valued function to which we will refer to $\varphi$ as ``potential defect''. Here $L\in \mathbb{R}^+$ stands for the width of the defect. Let $\{ \ell_j \in  \mathbb{R}^+ \}_{j\in\mathbb{Z}}$ be a sequence of non-negative numbers defined as follows,
\begin{equation}
\ell_{j} =  \left\{ \begin{array} 
            {r@{\quad \mbox{ if } \quad}l} 
   \delta_{j/2}   &  j \ \mbox{is even }  \\ 
   L  &   j\ \mbox{is odd},       \\ 
             \end{array} \right. 
\end{equation}
where $\{ \delta_j \in  \mathbb{R}^+ \}_{j\in\mathbb{Z}}$ is a set of independent and identically distributed random variables. Additionally let $L_n$ be defined as the partial sum of the $\ell_j$'s up to $n$ (we set $L_0 =0$),
\begin{equation}
L_n =  \left\{ \begin{array} 
            {r@{\quad \mbox{ if } \quad}l} 
   \sum_{j=0}^n  \ell_j &  n > 0  \\ 
   -\sum_{j=1}^{|n|}  \ell_j&   n<0,       \\ 
             \end{array} \right. 
\end{equation}

\begin{figure}[h]
\begin{center}
\scalebox{0.35}{\includegraphics{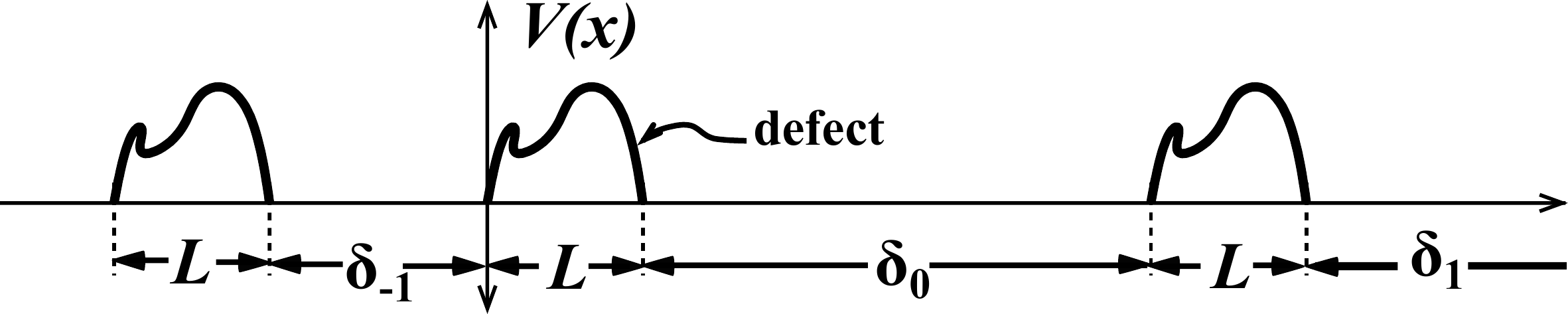}}
\end{center}
     \caption{
     Schematic representation of the potential model. 
              }
\label{fig:potential}
\end{figure}

Then, in terms of the above-defined quantities we can define the disordered potential $V(x)$ is defined as follows
\begin{equation}
\label{eq:random_potential}
V(x) =  \left\{ \begin{array} 
            {r@{\quad \mbox{ if } \quad}l} 
   \varphi(x-L_{2n})    &  L_{2n} \leq  x<L_{2n+1}    \\ 
   0  &   L_{2n+1} \leq  x<L_{2n+2}.       \\ 
             \end{array} \right. 
\end{equation}
This potential can be thought as a constant potential, $V(x)=0$, that has been ``contaminated'' with some defects, which are modeled through the potential profile $\varphi$. The distance between two consecutive defects is $\delta_j$, which is randomly chosen from  a prescribed distribution, while the width of the defect is a constant $L$. In Figure~\ref{fig:potential} we can appreciate an schematic representation of a realization of this potential. Additionally, notice that the equation of motion allows two types of motion, namely, running and locked trajectories. The condition to have running solutions is that the driving constant for ce $F$ in Eq.~\eqref{eq:motion} should be larger that the critical value $F_{\mathrm{c}}:=\sup_{x}\{|f(x)|\} = \sup_{x}\{|-\varphi^\prime(x)|\}$. On the other hand, to have locked trajectories we require that $F\leq F_{\mathrm{c}}$. In the following we will assume that the driving force $F$ is strictly above the critical value $F_{\mathrm{c}}$, i.e.,  $F > F_{\mathrm{c}} $, which means that every particle in the disordered potential moves always to the right and never gets stuck.

%%
%%==================== FIGURE =========================
%%
%%==================== FIGURE =========================
%

In the following we will assume that every random variable $\delta_j$ has a heavy tailed distribution. Particularly, we will chose a probability density function $\rho(x)$ given by,
\begin{equation} 
\label{heavy tailed dist}
\rho_\alpha (x) = \left\{ \begin{array} 
            {r@{\quad \mbox{ if } \quad}l} 
  \alpha  x^{-\alpha -1}   & x\geq 1   \\ 
   0  &   x < 1       \\ 
             \end{array} \right.  
\end{equation}

Notice that the larger distance between defects the fewer defects are present in the substrate. For $\alpha$ small, the distance between two successive defect are typically larger than the distance between two contiguous defects when $\alpha$ is large. Thus, we can interpret $\alpha$ as a parameter controlling the ``quantity'' of defects present in the substrate. 

We are interested in knowing the asymptotic behavior of typical trajectories $X_t$ for large $t$. To this end we will made use of generalized limit theorems~\cite{gnedenko1968limit,feller1960introduction,chazottes2015fluctuations} which have been shown to be useful in calculating the asymptotic behavior of typical trajectories in disordered media~\cite{kotulski1995asymptotic,salgado2013normal,salgado2014effective,salgado2015normal,salgado2015unbiased}. 

Notice that the equation of motion~\eqref{eq:motion} can be solved analytically on every ``piece'' of the potential. Indeed we can calculate the time that the particle spent in crossing every piece. First let us consider the time $\tau_{\mathrm{D}}$ that the particle takes to go across the defect. This quantity is given by,
\begin{equation}
\tau_{\mathrm{D}} = \int_0^L \frac{\gamma dx}{-\phi^\prime(x) + F}.
\end{equation}
On the other hand, the time $\tau_j$ that the particle takes to go from the $j$th defect to the $(j+1)$th one is given by 
\begin{equation}
\tau_j = \frac{\gamma }{F}\delta_j.
\end{equation}
Clearly, the time $\tau_j$ is a random variable that depends linearly on the (random) distance between two defects

Let us call ``unit cell'' the piece of the potential which contains a defect followed by the flat potential between such a defect and the next one. Then the total time $T_n$ that the particle takes to cross throughout the first $n$ unit cells is given by
\begin{equation}
\label{eq:T_n}
T_n = \sum_{j=0}^{n-1}\tau_j + n\tau_{\mathrm{D}} = \frac{\gamma}{F}\sum_{j=0}^{n-1}\delta_j  + n\tau_{\mathrm{D}}.
\end{equation}
Moreover, since the unit cells have a random length (given by $L + \delta_j$), the total displacement achieved by the particle during a time $T_n$ is given by,
\begin{equation}
X_n = \sum_{j=0}^{n-1}\delta_j + nL.
\end{equation}

The above definitions of $T_n$ and $X_n$ give us implicitly the (random) trajectory of a particle. However, we cannot extract directly from these expression how the mean displacement and the mean square displacement behaves as a function of time. First we need to perform an intermediate step. In order to have explicitly the  dependence of $X_n$ in terms of the time $T_n$ we will use the classical limit theorems for sums of random variables. We should notice that both, $X_n$ and $T_n$ are expressed in terms of the sum of independent and identically distributed random variables,
\begin{equation}
S_n = \sum_{j=0}^{n-1} \delta_j
\end{equation}
As we stated above, the random variables $\{ \delta_j\}_{ j\in\mathbb{N}}$ have heavy-tailed distributions, and therefore, the asymptotic properties of $S_n$ for large $n$ strongly depends on the exponent $\alpha$ of the distribution $\rho_\alpha$. As we will see below, we have several scenarios depending on the values of $\alpha$.

%%%%%%%%%%%%%%%%%%%%%%%%%%%%%%%%%%%%%%%%%%%%%%%%%%%%%%%%%%%%%%%%%%%%%%%%%%%%%%
%%%%%%%%%%%%%%%%%%%%%%%%%%%%%%%%%%%%%%%%%%%%%%%%%%%%%%%%%%%%%%%%%%%%%%%%%%%%%%
%
\section{Normal diffusion}
\label{sec:normal}
%
%%%%%%%%%%%%%%%%%%%%%%%%%%%%%%%%%%%%%%%%%%%%%%%%%%%%%%%%%%%%%%%%%%%%%%%%%%%%%%
%%%%%%%%%%%%%%%%%%%%%%%%%%%%%%%%%%%%%%%%%%%%%%%%%%%%%%%%%%%%%%%%%%%%%%%%%%%%%%

%%%%%%%%%%%%%%%%%%%%%%%%%%%%%%%%%%%%%%%%%%%%%%%%%%%%%%%%%%%%%%%%%%%%%%%%%%%%%%

\subsection{The case $\alpha >2$}

%%%%%%%%%%%%%%%%%%%%%%%%%%%%%%%%%%%%%%%%%%%%%%%%%%%%%%%%%%%%%%%%%%%%%%%%%%%%%%

We should remind that the distribution of $\delta_j$ have its first and second moments finite if $\alpha >2$. Since all the random variables $\{\delta_j\}_{j\in\mathbb{Z}}$ are assumed to be independent, it is clear that we can apply the central limit theorem~\cite{gnedenko1968limit,feller1960introduction}. Then, for sufficiently large $n$, this theorem implies the sum random variables $\sum_{j=0}^{n-1}\delta_j$ can be approximated by a normal random variable~\cite{gnedenko1968limit,feller1960introduction,salgado2013normal},
\begin{equation}
\sum_{j=0}^{n-1}\delta_j \approx n\bar{\delta} + \sqrt{n}\sigma_\delta Z ,
\end{equation}
where we defined $\bar{\delta}$ and $\sigma_\delta$ as the expected value and standard deviation of $\delta_j$ respectively,
\begin{eqnarray}
\bar{\delta} &:=& \mathbb{E}[\delta_j],
\\
\sigma_\delta &:=& \sqrt{\mbox{Var}(\delta_j)},
\end{eqnarray}
which in our case are explicitly given by
\begin{eqnarray} \label{eq:delta_mean}
\bar{\delta} &:=& \frac{\alpha}{\alpha- 1},
\\
\sigma_\delta^2 &:=&  \frac{\alpha}{\alpha -2} - \frac{\alpha^2}{(\alpha -1)^2}\label{eq:delta_var}.
\end{eqnarray}

Within this approximation we can rewrite the time $T_n$ and the displacement $X_n$, for asymptotically large $n$, as
\begin{eqnarray}
T_n &\approx& n\tau_{\mathrm{D}} + n \frac{\gamma \bar{\delta} }{F} + \frac{\gamma \sigma_\delta  }{F}\sqrt{n} Z, 
\label{eq:Tn}
\\
X_n &\approx& n L + n\bar{\delta} + \sqrt{n}\sigma_\delta Z.
\label{NPosition}
\end{eqnarray}

Calling $N_t$ the number of unit cells that the particle has crossed during a time $t$, we can define implicitly $N_t$ by the equation $T_{N_{t}}=t$, as it has been done in~\cite{salgado2013normal}. From Eq.~\eqref{eq:Tn} we can observe that the random variable $N_t$ is related to the random variable $Z$ as follows,
\begin{equation}
\frac{t-N_t(\tau_{\mathrm{D}}+\frac{\gamma \bar{\delta}}{F})}{\sqrt{\frac{\gamma^{2}}{F^{2}}\sigma_\delta^{2} N_t}}\simeq Z.
\end{equation}

The most probable values for $Z$ are around zero, which implies that the distribution of $N_t$ should be centered around the root of a function,
\[
\psi(N_t):=\frac{t-N_t(\tau_{\mathrm{D}}+\frac{\gamma \bar{\delta}}{F})}{\sqrt{\frac{\gamma^{2}}{F^{2}}\sigma_\delta^{2} N_t}}.
\]
We can see that $t/({\tau_{\mathrm{D}}+\frac{\gamma \bar{\delta}}{F}})$ is the unique root of $\psi(N_t)$. To find an expression of $N_t$ in terms of $Z$ we  proceed to make a linear expansion of $	\psi(N_t) $ around its root. We have
\begin{equation}
\psi(N_t)\simeq -\frac{(\tau_{\mathrm{D}}+\frac{\gamma \bar{\delta}}{F})^{\frac{3}{2}}}{\frac{\gamma}{F}\sigma_\delta \sqrt{ t}}(N_t - \frac{t}{\tau_{\mathrm{D}}+\frac{\gamma \bar{\delta}}{F}})+O(t^{-\frac{3}{2}})
\end{equation}
from above we obtain
\begin{equation}
-\frac{(\tau_{\mathrm{D}}+\frac{\gamma \bar{\delta}}{F})^{\frac{3}{2}}}{\frac{\gamma}{\delta}\sigma_\delta \sqrt{t}}(N_t - \frac{t}{\tau_{\mathrm{D}}+\frac{\gamma \bar{\delta}}{F}}) \approx Z.
\end{equation}
 Finding the value for $N_t$, this can be expressed  by   
\begin{equation}
\label{Ncells}
N_t \approx \frac{t}{\tau_{\mathrm{D}} + \frac{\gamma \bar{\delta}}{F}} - \frac{\frac{\gamma}{F}\sigma_\delta \sqrt{t}Z}{(\tau_{\mathrm{D}}+\frac{\gamma \bar{\delta}}{F})^{\frac{3}{2}}}.
\end{equation}
Now we can substitute Eq.~\eqref{Ncells}  into Eq.~\eqref{NPosition} to find the expected value and  variance for the particle position $X_t$. This gives
\begin{eqnarray}
\mathbb{E}[X_t] &=& \frac{L+\bar{\delta}}{\tau_{\mathrm{D}}+\frac{\gamma \bar{\delta}}{F}}t,
\\
\mbox{Var}(X_t) &=& \Bigg(\frac{1}{\sqrt{\tau_{\mathrm{D}}+\frac{\gamma \bar{\delta}}{F}}}-\frac{\frac{\gamma}{F}(L+\bar{\delta})}{(\tau_{\mathrm{D}}+\frac{\gamma \bar{\delta}}{F})^{\frac{3}{2}}}\Bigg)^{2}\sigma_\delta^{2}t.
\end{eqnarray}
Applying the usual definitions of particle current, $J_{\mathrm{eff}}$, and effective diffusion coefficient, $D_{\mathrm{eff}}$, we obtain the following expressions,
\begin{eqnarray}\label{J_Diff_Norm}
J_{\mathrm{eff}} &=& \lim_{t\to\infty} \frac{\mathbb{E}[X_{t}]}{t} = \frac{L+\bar{\delta}}{\tau_{\mathrm{D}}+\frac{\gamma \bar{\delta}}{F}},
\\
D_{\mathrm{eff}} &=& \lim_{t\to\infty} \frac{\mbox{Var}(X_t)}{2t}
\nonumber
\\ 
\label{D_Diff_Norm}
&=& \Bigg(\frac{1}{\sqrt{\tau_{\mathrm{D}}+\frac{\gamma \bar{\delta}}{F}}}-\frac{\frac{\gamma}{F}(L+\bar{\delta})}{(\tau_{\mathrm{D}}+\frac{\gamma \bar{\delta}}{F})^{\frac{3}{2}}}\Bigg)^{2} \frac{\sigma_\delta^{2}}{2}.
\end{eqnarray}

%%%%%%%%%%%%%%%%%%%%%%%%%%%%%%%%%%%%%%%%%%%%%%%%%%%%%%%%%%%%%%%%%%%%%%%%%%%%%%
%%%%%%%%%%%%%%%%%%%%%%%%%%%%%%%%%%%%%%%%%%%%%%%%%%%%%%%%%%%%%%%%%%%%%%%%%%%%%%
%
\section{Anomalous diffusion}
\label{sec:anomalous}
%
%%%%%%%%%%%%%%%%%%%%%%%%%%%%%%%%%%%%%%%%%%%%%%%%%%%%%%%%%%%%%%%%%%%%%%%%%%%%%%
%%%%%%%%%%%%%%%%%%%%%%%%%%%%%%%%%%%%%%%%%%%%%%%%%%%%%%%%%%%%%%%%%%%%%%%%%%%%%%

%%%%%%%%%%%%%%%%%%%%%%%%%%%%%%%%%%%%%%%%%%%%%%%%%%%%%%%%%%%%%%%%%%%%%%%%%%%%%%

\subsection{The case $ 1 <\alpha <2 $}

%%%%%%%%%%%%%%%%%%%%%%%%%%%%%%%%%%%%%%%%%%%%%%%%%%%%%%%%%%%%%%%%%%%%%%%%%%%%%%

If the distribution of $\delta_j$ (given by Eq.~\eqref{heavy tailed dist}) is such that the exponent $\alpha$ is in the interval $ 1 < \alpha < 2$ we have that the variance of $\delta_j$ is longer finite but its mean remains finite. In that case, the central limit theorem cannot be applied in its standard form.  Actually, the sum $\sum_{j=0}^{n-1}\delta_j $ tends to a stable law if such a sum is adequately normalized. According to a well-known theorems in probability~\cite{gnedenko1968limit,feller1960introduction}, we have that 
\begin{equation}
\frac{\sum_{j=0}^{n-1}\delta_j - n\bar{\delta} }{ n^{1/\alpha}} \to W.
\end{equation}
where $W$ is a $\alpha$-stable random variable. This means that we can approximate the sum $\sum_{j=0}^{n-1}\delta_j $ by 
\begin{equation}
\label{eq:sum_delta_2}
\sum_{j=0}^{n-1}\delta_j  \approx n \bar{\delta}  + n^{1/\alpha} W.
\end{equation}
Now, we proceed to define the random variable $N_t$ by using the relationship $T_{N_t}=t$.  Recalling the definition for $T_n$ given in Eq.~\eqref{eq:T_n} we obtain that $N_t$ satisfies the equation,
\begin{equation}
\frac{\gamma }{ F } \big(N_t \bar{\delta}  + N_t^{1/\alpha} W \big) + N_t \tau_{\mathrm{D}} \approx t.
\end{equation} 
The above equation implicitly defines a transformation from the random variable $W$ to $N_t$. Notice that the above equation can be rewritten as
\begin{equation}
\label{eq:psi_1}
\psi (N_t) := \frac{t-\tau_{\mathrm{c}} N_t}{\frac{\gamma}{F} N_t^{1/\alpha}}  \approx W, 
\end{equation} 
where we have denoted by $\tau_{\mathrm{c}}$ the quantity
\begin{equation}
\label{eq:def_tau_c}
\tau_{\mathrm{c}} := \frac{\gamma \bar{\delta}}{F} + \tau_{\mathrm{D}}.
\end{equation}

Eq.~\eqref{eq:psi_1} means that the transformation from $W$ to $N_t$ is mediated by the inverse function $\psi^{-1}$. As it has been shown in Ref.~\cite{salgado2013normal} we have that the asymptotic behavior of $N_t$ for $t\to \infty$ is given by, 
\begin{equation}
\label{eq:Nt_2}
N_t \approx \frac{1}{\tau_{\mathrm{c}} } t + \frac{\frac{\gamma}{F} \, t^{1/\alpha}}{\tau_{\mathrm{c}}^{1+1/\alpha}}\, W.
\end{equation}

Now, in order to have an expression for the displacement of the particle as a function of $t$, we will use the approximation given in Eq.~\eqref{eq:sum_delta_2} to obtain an approximation of $X_n$ for $n \to \infty$. Recalling that  $X_n$ is given by
\[
X_n = \sum_{j=0}^{n-1}\delta_j + nL,
\]
we can observe that 
\begin{equation}
X_n \approx n \bar{\delta}  + n^{1/\alpha} W + nL.
\end{equation}
Next, if we substitute $n=N_t$ , given in Eq.~\eqref{eq:Nt_2} into the above expression for $X_n$ we obtain,
\begin{eqnarray}
X_t &\approx& ( L+\bar{ \delta}) N_t + N_t^{1/\alpha} W,
\nonumber
\\
&\approx& ( L+\bar{ \delta}) \left( \frac{t}{ \tau_{\mathrm{c}} }  + \frac{ \frac{\gamma}{F} \, t^{1/\alpha}}{\tau_{\mathrm{c}}^{1+1/\alpha}}\, W \right) 
\nonumber
\\
&+&  \bigg(\frac{t}{ \tau_{\mathrm{c}} }  + \frac{\frac{\gamma}{F} \, t^{1/\alpha}}{\tau_{\mathrm{c}}^{1+1/\alpha}}\, W \bigg)^{1/\alpha} W.
\end{eqnarray}
Now, if we retain the leading terms in the above expression we have that,
\begin{eqnarray}
X_t &\approx& 
\bigg(\frac{L+\bar{ \delta}}{ \tau_{\mathrm{c}} }\bigg)\,  t 
+
 \bigg(  \frac{\frac{\gamma}{F}  ( L+\bar{ \delta}) }{\tau_{\mathrm{c}}^{1+1/\alpha}} + \frac{1}{\tau_{\mathrm{c}}^{1/\alpha} }  \bigg)\, t^{1/\alpha}\, W
 \nonumber
 \\
 %&+& \frac{L+\bar{ \delta}}{ \tau_{\mathrm{c}} } t 
 &+& O(t^{2/\alpha -1})
\end{eqnarray}

With the above result we can see that the mean displacement of an ensemble of particles is given by
\begin{equation}
\mathbb{E}[X_t] \approx \frac{L+\bar{ \delta}}{ \frac{\gamma}{F} + \tau_\mathrm{D}}\, t, \qquad \mbox{for} \, \, t\to\infty.
\end{equation}
On the other hand, the square fluctuations of $X_t$ grow as
\begin{equation}
\label{eq:SD_12}
(X_t - \mathbb{E}[X_t])^2 \approx\bigg(  \frac{\frac{\gamma}{F}  ( L+\bar{ \delta}) }{\tau_{\mathrm{c}}^{1+1/\alpha}} + \frac{1}{\tau_{\mathrm{c}}^{1/\alpha} }  \bigg)^2 t^{2/\alpha} W^2,
\end{equation}
which means that the diffusion exponent $\beta$ is given by,
\begin{equation}\label{eq:superdif}
\beta = \frac{2}{\alpha},
\end{equation}
for $1 < \alpha <2 $. In this case we see clearly that the system undergoes a transition from normal to anomalous superdiffusion when the parameter $\alpha$ diminishes. In this anomalous phase, the mean displacement is still finite, and therefore the particle current can be written as
\begin{equation}
J_{\mathrm{eff}} = \frac{L+\bar{ \delta}}{\tau_\mathrm{c}}.
\end{equation}

%%%%%%%%%%%%%%%%%%%%%%%%%%%%%%%%%%%%%%%%%%%%%%%%%%%%%%%%%%%%%%%%%%%%%%%%%%%%%%

\subsection{The case $ 0 <\alpha < 1 $}

%%%%%%%%%%%%%%%%%%%%%%%%%%%%%%%%%%%%%%%%%%%%%%%%%%%%%%%%%%%%%%%%%%%%%%%%%%%%%%

Now we will explore the case in which the parameter $\alpha$ is in the range $0 < \alpha < 1$. In this case we have that the mean value of $\delta_j$ diverge, which means that the approximation for the sum of $\delta_j$ given in Eq.~\eqref{eq:sum_delta_2} cannot be applied. However the sum  $\sum_{j=0}^{n-1} \delta_j $ still converge to a stable law if it is normalized appropriately. Explicitly we have that~\cite{gnedenko1968limit,feller1960introduction},
\begin{equation}
\frac{\sum_{j=0}^{n-1}\delta_j }{ n^{1/\alpha}} \to W,
\end{equation}
where $W$ has an $\alpha$-stable distribution~\cite{feller1960introduction}. In this case we can approximate the sum of random variables by
\begin{equation}
\label{eq:sum_delta_3}
\sum_{j=0}^{n-1}\delta_j  \approx n^{1/\alpha} W,
\end{equation}
which allows us to obtain an asymptotic expression for $N_t$ by using the relationship $T_{N_t} = t$, 
\begin{equation}
\frac{\gamma }{ F }  N_t^{1/\alpha} W  + N_t \tau_{\mathrm{D}} \approx t.
\end{equation} 
Now we proceed as in the above cases, i.e., we will obtain an asymptotic expression for $N_t$ for $t\to \infty$. For such purpose we first write the above equation as follows
\[
N_t \approx \frac{t^\alpha}{ \left(\frac{\gamma}{F} W \right)^\alpha}\bigg( 1 - \frac{\tau_{\mathrm{D} } N_t}{t} \bigg)^\alpha,
\]
and then we use this expression recursively in order to obtain an asymptotic expression for $t\to \infty$. We obtain
\begin{equation}
N_t\approx \frac{t^\alpha}{ \left(\frac{\gamma}{F} W  \right)^\alpha} + O(t^{2\alpha -1}) \qquad \mbox{for} \, \, t\to \infty.
\end{equation}

Now, in order to see how $X_t $ behaves in time, we use the approximation~\eqref{eq:sum_delta_3} to obtain an asymptotic expression for $X_n$ for large $n$,
\begin{equation}
X_n = \sum_{j=0}^{n-1}\delta_j + nL \approx n^{1/\alpha} W + nL.
\end{equation}
Thus, if we substitute $n$ by $N_t$ into the above equation we obtain
\begin{eqnarray}
X_t &\approx& \bigg( \frac{t^\alpha}{ \left(\frac{\gamma}{F} W  \right)^\alpha}  \bigg)^{1/\alpha} W +  \frac{ L t^\alpha}{ \left(\frac{\gamma}{F} W  \right)^\alpha}
\nonumber
\\
&=&
\frac{F}{\gamma} \, t + \frac{1}{  \left(\frac{\gamma}{F} W  \right)^\alpha  }\, t^\alpha.
\label{eq:xt_3}
\end{eqnarray}

The above result for $X_t$ allows us to obtain an expression for the particle current. First notice that the expected value of $X_t$ is give by,
\[
\mathbb{E}[X_t] = \frac{F}{\gamma} \, t + \frac{ t^\alpha}{  \left(\frac{\gamma}{F}  \right)^\alpha  }\, \mathbb{E}[W^{-\alpha}],
\]
from which, after noticing that $\mathbb{E}[W^{-\alpha}]$ is finite and recalling that $0<\alpha <1$, we obtain,
\begin{equation}
J_{\mathrm{eff}} = \lim_{t\to \infty} \frac{\mathbb{E}[X_t]}{t} = \frac{F}{\gamma}.
\end{equation}

The expression for $X_t$ that we obtained in Eq.~\eqref{eq:xt_3} also allows us to calculate the asymptotic behavior of the diffusion. Indeed we have that
\begin{eqnarray}
\label{eq:SD_01}
\mbox{Var}(X_t) &:=& \mathbb{E}[(X_t - \mathbb{E}[X_t])^2] 
\nonumber
\\
&=& \bigg(\frac{ t^\alpha}{  \left(\frac{\gamma}{F}  \right)^\alpha  }\, \bigg)^2 \mathbb{E}[(  W^{-\alpha} - \mathbb{E}[W^{-\alpha}] )^2] 
\end{eqnarray}
which means that the means square displacement of the particle distribution grows in time as $t^{2\alpha}$, giving a diffusion exponent 
\begin{equation}\label{eq:subdif}
\beta = 2\alpha.
\end{equation}
The above result implies that the anomalous regime for $0< \alpha <1$ exhibits two different behaviors, namely, an anomalous superdiffusive phase for $1/2 <\alpha < 1$, and an anomalous subdiffusive phase for $0 < \alpha <1/2 $. Then, a transition from superdiffusion to subdiffusion occurs at the critical value $\alpha  = 1/2$.

%%%%%%%%%%%%%%%%%%%%%%%%%%%%%%%%%%%%%%%%%%%%%%%%%%%%%%%%%%%%%%%%%%%%%%%%%%%%%%

\subsection{The marginal cases $\alpha = 2$ and $\alpha = 1$}

%%%%%%%%%%%%%%%%%%%%%%%%%%%%%%%%%%%%%%%%%%%%%%%%%%%%%%%%%%%%%%%%%%%%%%%%%%%%%%

In this section we will explore the transport properties for the special values $\alpha = 2$ and $\alpha =1$. We should emphasize that in these cases the asymptotic behavior of $S_n$ for large $n$ is rather different than in the above cases. First lets us consider the marginal value $\alpha = 2$. In this case the variance of $\delta_j$ diverge. However, according to Ref.~\cite{feller1960introduction} we have that $\sum_{j=0}^{n-1} \delta_j$ still converge to a normal distribution if it is appropriately normalized. Indeed we have that~\cite{gnedenko1968limit,feller1960introduction}
\[
\frac{\sum_{j=0}^{n-1} \delta_j - n \bar{\delta}}{\sqrt{n \ln n}}\to W, \qquad \mbox{for} \, \, n\to\infty,
\]
where $W$ is a normal random variable. The above means that if $n$ is large enough we can approximate the sum $\sum_{j=0}^{n-1} \delta_j$ as follows,
\begin{equation}
\label{eq:sum_delta_4}
\sum_{j=0}^{n-1} \delta_j \approx  n \bar{\delta} + \sqrt{n \ln n} \, W.
\end{equation}
As above, we use the equation $T_{N_t} = t$ to approximate $N_t$ for large $t$. After some calculations we obtain that $N_t$ satisfy the equation,
\begin{equation} 
\label{eq:psi_4}
\psi(N_t) := \frac{ t- \tau_{\mathrm{c}} N_t}{ \left( N_t \ln (N_t) \right)^{1/2}} \approx \frac{\gamma}{F} \, W.
\end{equation}
As we argued in preceding sections, the random variable $ \frac{\gamma}{F} \, W $ has zero mean value, which implies that the most probable values of $N_t$ are around the (unique) root of $\psi(N_t)$. If we expand $\psi$ around $N_t = t/\tau_{\mathrm{c}}$ we obtain,
\begin{equation}
\psi(N_t) \approx - \frac{\tau_{\mathrm{c}} }{ \left[ \frac{t}{ \tau_{\mathrm{c}} }\ln\left( \frac{t}{ \tau_{\mathrm{c}}} \right)  \right]^{1/2}}\, \left( N_t -\frac{t}{\tau_{\mathrm{c}}} \right),
\end{equation}
which allows us to obtain $N_t$ in terms of $W$ by means of Eq.~\eqref{eq:psi_4}, 
\begin{equation}
\label{eq:N_t_4}
N_t \approx \frac{t}{ \tau_{\mathrm{c}} } - 
\frac{ 1 }{\tau_{\mathrm{c}} }
 \left[ \frac{t}{ \tau_{\mathrm{c}} }\ln\left( \frac{t}{\tau_{\mathrm{c}} } \right)  \right]^{1/2}\, W.
\end{equation}

Now, we use again the approximation for the sum given in~\eqref{eq:sum_delta_4} to obtain an asymptotic expression for $X_n$. This gives,
\begin{equation}
X_n \approx n \bar{\delta} + \sqrt{n \ln n} \, W + nL.
\end{equation}
Next we substitute $n$ by $N_t$ in the above expression, resulting in an expression for the displacement $X_t$ given by,
\begin{eqnarray}
X_n &\approx & \left( \frac{ L+ \bar{\delta}}{\tau_{\mathrm{c}}  } \right) \, t
\nonumber
\\
 &+& 
\left( \frac{F}{\gamma} - \frac{L+ \bar{\delta} }{ \tau_{\mathrm{c}}  }  \right) 
\bigg[ \left(\frac{t}{\tau_{\mathrm{c}}  } \right) \ln \left(\frac{t}{\tau_{\mathrm{c}} } \right)\bigg]^{1/2} \, W.
\nonumber
\end{eqnarray}

Thus, the last expression implies that mean displacement of the trajectory grows linearly in time. This means that the particle current is well defined and has the value
\begin{equation}
J_{\mathrm{eff}} = \frac{ L+\bar{\delta} }{\tau_{\mathrm{c}}  } .
\end{equation}
On the other hand, the mean square fluctuations of the trajectory can also be calculates, giving,
\begin{equation}
\mbox{Var}(X_t) = \left( \frac{F}{\gamma} + \frac{L+ \bar{\delta} }{ \tau_{\mathrm{c}}  }  \right)^2 
\left(\frac{t}{\tau_{\mathrm{c}}  } \right) \ln \left(\frac{t}{\tau_{\mathrm{c}} } \right)  \, \mathbb{E}[W^2].
\end{equation}
The above result states that the mean square fluctuations do not grow linearly in time nor as a power law, but it still grows faster than linear by the presence of the logarithm term $\ln(t/\tau_{\mathrm{c}})$. This kind of behavior of the mean square displacement is commonly called \emph{marginal superdiffusion}. 

Another case that it is necessary to explore separately is $\alpha =1$. In this case the sum $S_n$ converge to a stable law, but the renormalizing  factor is not of the form $n^{1/\alpha}$. Actually we have that~\cite{feller1960introduction},
\begin{equation}
\frac{\sum_{j=0}^{n-1} \delta_j }{n\ln( n)} \to W,\qquad \mbox{for} \,\, n\to \infty,
\end{equation}
where $W$ is a random variable with a $\alpha$-stable distribution with $\alpha =1$. In this case we approximate the sum $S_n$ by 
\[
\sum_{j=0}^{n-1} \delta_j  \approx n\ln (n) W,
\]
which allows us to estimate the asymptotic behavior of $N_t$ by the equation,
\begin{equation}
\label{eq:N_t_5}
\frac{\gamma}{F} N_t\ln (N_t) W + N_t \tau_{\mathrm{D}} \approx t.
\end{equation}

Following a similar reasoning as in preceding sections we use the above relation to obtain an asymptotic expression for $N_t$. Indeed, if we notice that the leading term in Eq.~\eqref{eq:N_t_5} is $N_t \ln(N_t)$ we obtain,
\begin{equation}
N_t\approx  \frac{t/\left(\frac{\gamma}{F}W\right)}{\ln\left[ t/\left(\frac{\gamma}{F}W\right)  \right]}.
\end{equation}
Next, if we substitute $n$ by $N_t$ into the expression for $X_n$ we can observe that,
\begin{equation}
X_t \approx \frac{F}{\gamma}\, t + \left(L-\frac{F\tau_{\mathrm{D}} }{\gamma} \right)\frac{t/\left(\frac{\gamma}{F}W\right)}{\ln\left[ t/\left(\frac{\gamma}{F}W\right)  \right]}.
\end{equation}
The last result means that the mean displacement exists and grows linearly in time, which gives for the particle current,
\[
J_{\mathrm{eff}} = \frac{F}{\gamma}.
\]
Additionally we obtain that, for $\alpha = 2$ the square displacement grows as 
\[
\left( X_t - \mathbb{E}[X_t]  \right)^2 \sim \left( \frac{t}{\ln(t)}\right)^2
\]
which can be considered as \emph{marginally ballistic} since the square displacement grows nearly as $t^2$ but this growth is screened by the inverse logarithmic factor.

Finally let us summarize the asymptotic behavior of the mean displacement and the square displacement $\left( X_t - \mathbb{E}[X_t]\right)^2$ for different values of $\alpha$. We obtain that the particle current is given by 
\begin{equation} 	\label{eq:Jef}
J_{\mathrm{eff}} =   \left\{ \begin{array} 
            {r@{\quad \mbox{ if } \quad}l} 
   F/\gamma    &   0 < \alpha \leq  1     \\ 
  \frac{ L + \bar{\delta}}{ \tau_{\mathrm{D}} + \frac{\gamma \bar{\delta} }{ F}}   &  \alpha >1  .  \\    
             \end{array} \right. 
\end{equation}
On the other hand, the square displacement has the asymptotic behavior,
\begin{equation}
\label{eq:MSD}
\left(X_t - \mathbb{E}[X_t] \right)^2\sim  \left\{ \begin{array} 
            {r@{\quad \mbox{ if } \quad}l} 
  t^{2 \alpha}   &   0 < \alpha <  1     \\
  t^2/\ln^2(t)   &   \alpha =   1     \\
  t^{2/\alpha}   &   1 < \alpha <  2     \\
  t\ln(t)        &  \alpha = 2    \\
  t              &  \alpha >2.
             \end{array} \right. 
\end{equation}

%%%%%%%%%%%%%%%%%%%%%%%%%%%%%%%%%%%%%%%%%%%%%%%%%%%%%%%%%%%%%%%%%%%%%%%%%%%%%%
%%%%%%%%%%%%%%%%%%%%%%%%%%%%%%%%%%%%%%%%%%%%%%%%%%%%%%%%%%%%%%%%%%%%%%%%%%%%%%
%
\section{Numerical simulations}
\label{sec:simulations}
%
%%%%%%%%%%%%%%%%%%%%%%%%%%%%%%%%%%%%%%%%%%%%%%%%%%%%%%%%%%%%%%%%%%%%%%%%%%%%%%
%%%%%%%%%%%%%%%%%%%%%%%%%%%%%%%%%%%%%%%%%%%%%%%%%%%%%%%%%%%%%%%%%%%%%%%%%%%%%%

In order to test the theoretical results presented in Section~\ref{sec:anomalous}, we perform numerical simulations of our model.  We numerically solve the equation of motion given in Eq.~\eqref{eq:motion} for an ensemble of  non-interacting particles. Different  particles are placed in different realizations of the random potential described by Eqs.~\eqref{eq:random_potential} and~\eqref{heavy tailed dist}. Thus, once  we  have obtained the time series for the particle position, we compute the mean displacement and the square displacement by averaging over all the time series obtained. This actually corresponds to average over the ensemble of random potentials. For the sake of simplicity we model the defects of the random potential model by means of symmetric peaks with constant height and width. Thus the potential profile modeling the defects is defined as
\begin{equation}
\varphi(x) =  \left\{ \begin{array} 
            {r@{\quad \mbox{ if } \quad}l} 
  2 a x /L     &   0<x<\frac{L}{2}     \\ 
   2a(L-x)/L    &  \frac{L}{2}<x<L ,    \\ 
             \end{array} \right. 
\end{equation}
where $a$ and $L$ stand for the height and width of the  ``potential peak''. The corresponding random force field is given by,
\begin{equation}
-\varphi^{\prime}(x) =  \left\{ \begin{array} 
            {r@{\quad \mbox{ if } \quad}l} 
   -2 a /L    &  0<x<\frac{L}{2}     \\ 
    2 a  /L    &  \frac{L}{2}<x<L     \\ 
             \end{array} \right. 
\end{equation}
%%
%%==================== FIGURE =========================
%%
\begin{figure}[h]
\begin{center}
\scalebox{0.35}{\includegraphics{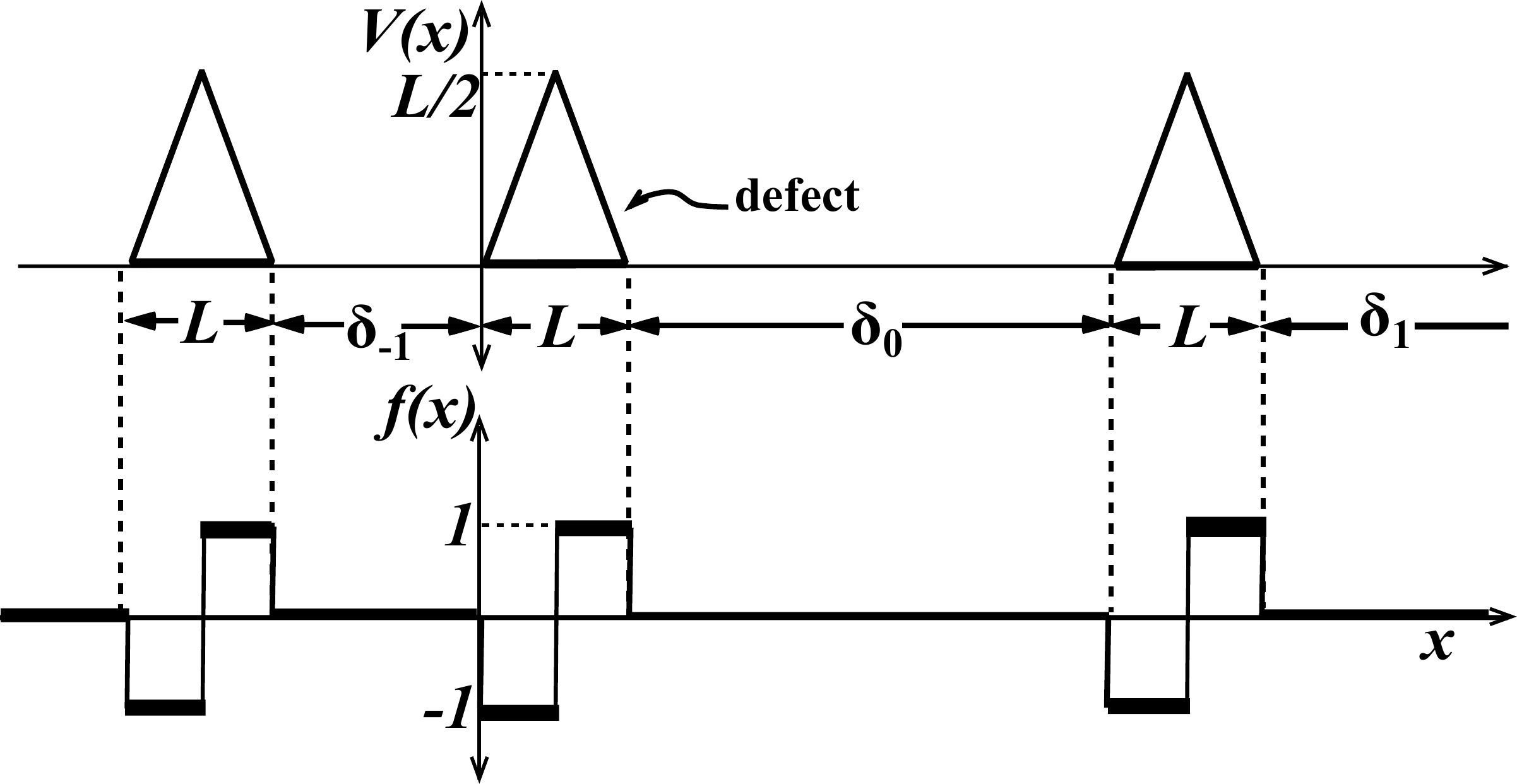}}
\end{center}
     \caption{
     Schematic representation of the random potential and the corresponding random force field. The height and width of the potential peak are $a$ and $L$ respectively. The distance between $j$th and the $(j+1)$th peaks is a random variable  $\delta_j$ whose distribution is given in Eq.~\eqref{heavy tailed dist}.
              }
\label{fig:potential simulations}
\end{figure}
%%==================== FIGURE =========================
%

In the numerical simulations we fixed the parameter values $L=1$ and $a = 1/2$. We have also taken the driving force $F$ above the critical one, which, according to our random potential model, is given by 
\[
F_{\mathrm{c}}  =\max_x |-\varphi^\prime(x)|= 2a/L = 1.
\] 
This choice for the driving force ensures the absence of locked trajectories.

\begin{figure}[ht]
\begin{center}
\scalebox{0.25}{\includegraphics{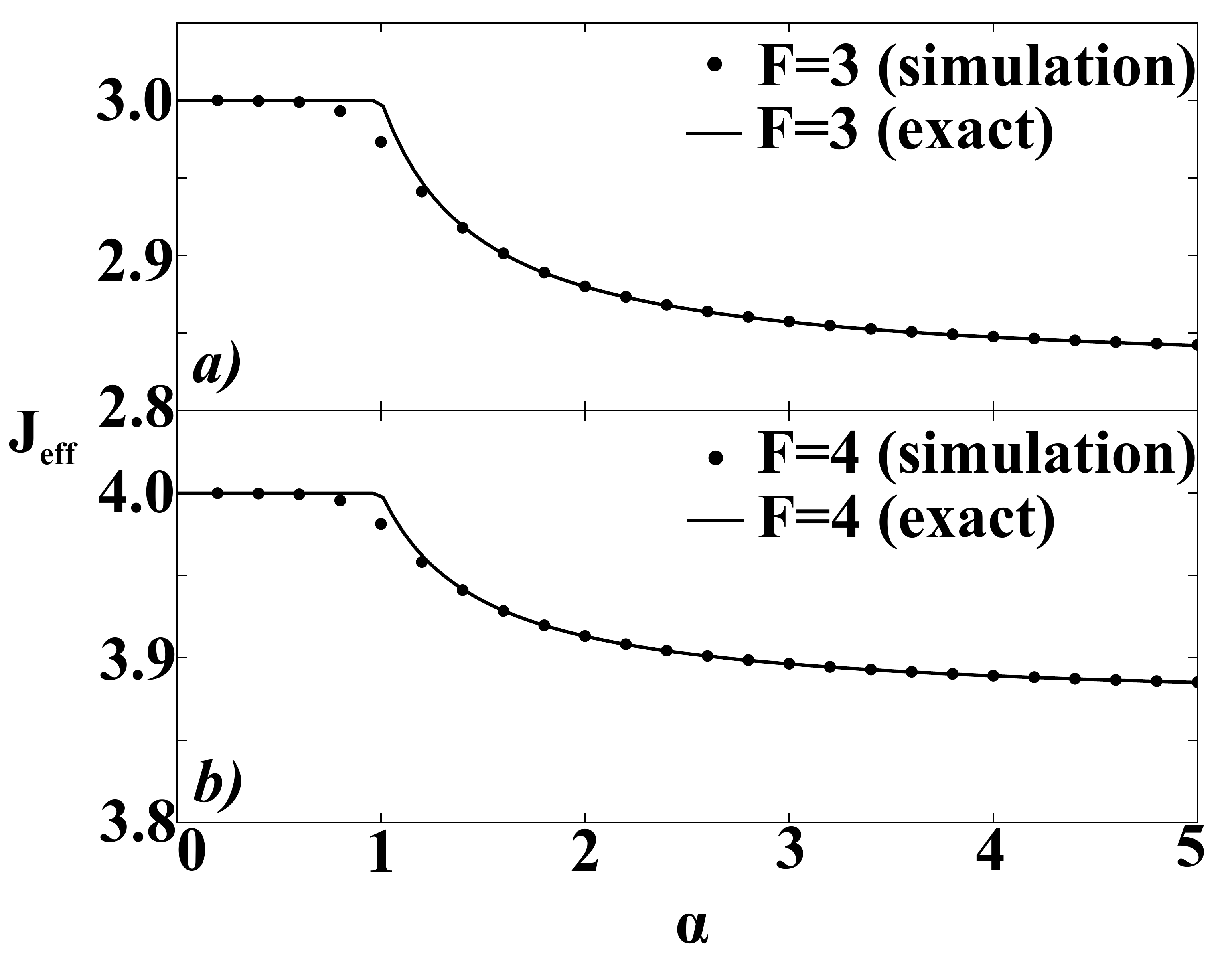}}
\end{center}
     \caption{
     Particle current as a function of $\alpha$. (a) We plot the particle current, by using the exact result given in Eq.~\eqref{eq:Jef}, for $F=3$ (solid line). We also show the particle current obtained from the numerical simulation of 500 particles during a time of $10^5$ arbitrary units, for $F=3$ (filled circles). (b) As in (a) but using $F=4$. This graph allows us to appreciate the abrupt change in the behavior of $J_{\mathrm{eff}}$ as $\alpha$ decreases.                    
      }
\label{fig:Jeff_alpha}
\end{figure}

In Fig.~\ref{fig:Jeff_alpha} we show the particle current as a function of the parameter $\alpha$ obtained by using the exact formula~\eqref{eq:Jef}. We plot the particle current for two different values of the driving force, namely $F=3$ (solid line) and $F=4$  (dashed line). We also plot the particle current obtained from numerical simulation for the same values of the parameter, i.e., for $F=3$ (filled circles) and $F=4$ (open squares). The numerical simulation were performed as follows. 
We solved the equation of motion, Eq.~\eqref{eq:motion}, for $500$ particles placed on random potentials during  a time of $10^5$ arb.~units. According to our simulation scheme, different particles move on different random potentials.  After that, we obtained the mean displacement the corresponding quantities over the ensemble of trajectories obtained from the simulations. 

It is important to emphasize that the  system undergoes a kind of second order ``phase transition'' in the sense that the particle current  curve changes continuously with the parameter $\alpha$, but its derivative does not. Indeed we observe that in the range $0 < \alpha \leq 1$ the particle current remains constant, a phenomenon which seems counterintuitive. This is because the presence of the defects has the effect of delaying the particles. Thus we would expect that if $\alpha$ increases then the particle current diminishes. As $\alpha$ increases beyond the critical value $\alpha = 1$, the particle current starts decreasing due to the presence of the defects as expected from the above reasoning.  

%%
%%==================== FIGURE =========================
%%
%%==================== FIGURE =========================
%

\begin{figure}[h]
\begin{center}
\scalebox{0.25}{\includegraphics{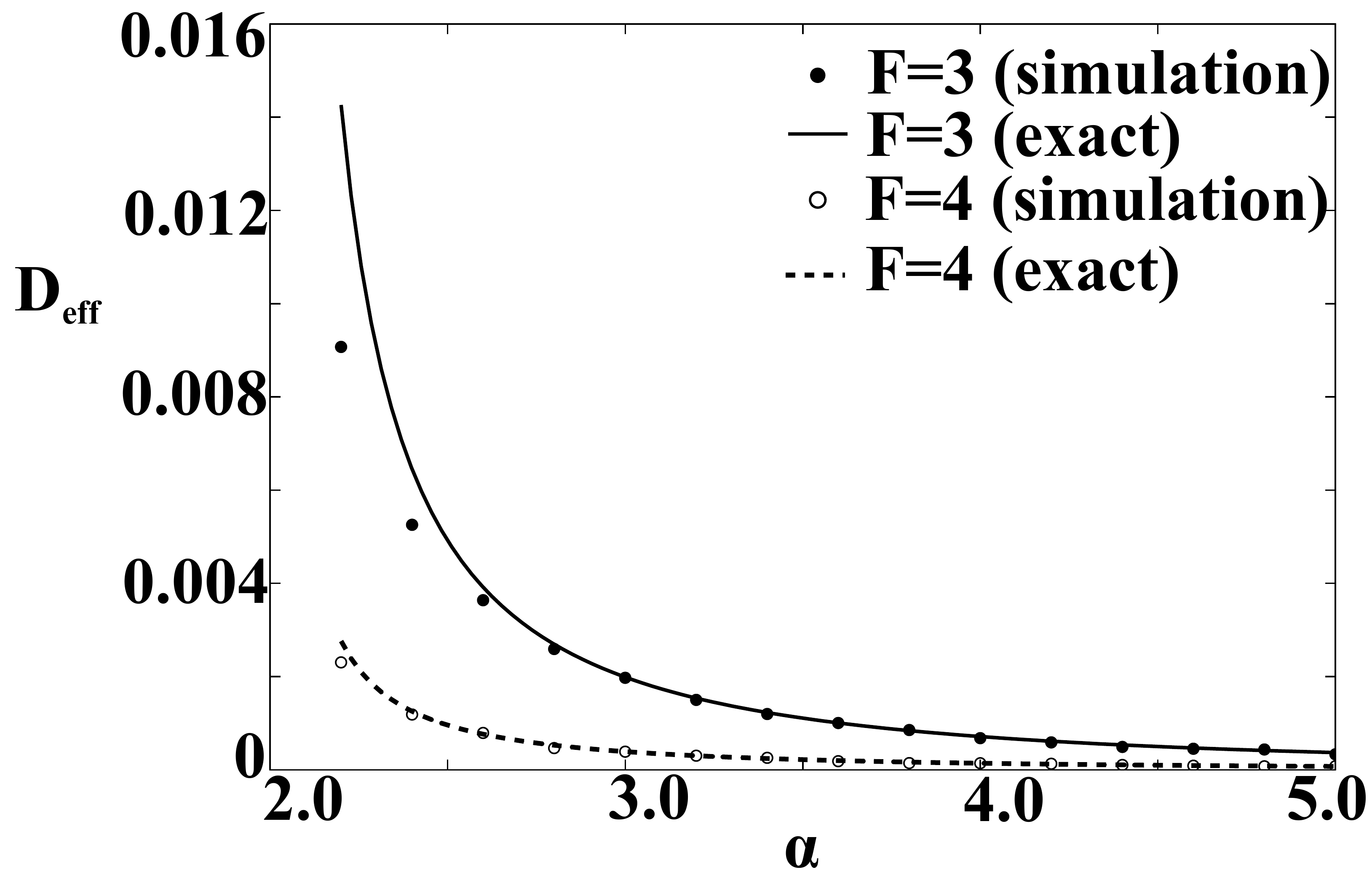}}
\end{center}
     \caption{
     Effective diffusion coefficient as a function of $\alpha$. We plot the effective diffusion coefficient by using the exact formula  given in Eq.~\eqref{D_Diff_Norm}, for two different values of the driving force $F=3$ (solid line) and $F=4$ (dashed line). We also show the diffusion coefficient obtained from the numerical simulation of 500 particles during a time of $10^5$ arbitrary units, for $F=3$ (filled circles) and $F=4$ (open  circle) for several values of $\alpha$.       
      }
\label{fig:Deff_alpha}
\end{figure}

In Fig.~\ref{fig:Deff_alpha} we show the behavior of $D_{\mathrm{eff}}$ as a function of $\alpha$ for two different values of the driving force. We use the the exact result given in Eq.~\eqref{D_Diff_Norm}. It is clear that this expression is only valid for $\alpha > 2$ because in this case the diffusion is normal. We plot the theoretical curves for the cases $F=3$ (solid line)  and $F=4$ (dashed line), which are compared with the corresponding coefficients obtained from numerical simulations. The simulations were performed by numerically solving the equation of motion~\eqref{eq:motion} for $500$ particles placed on random potentials during a time of $10^5$ arb.~units.

%%
%%==================== FIGURE =========================
%%
%%==================== FIGURE =========================
%

\begin{figure}[ht]
\begin{center}
\scalebox{0.23}{\includegraphics{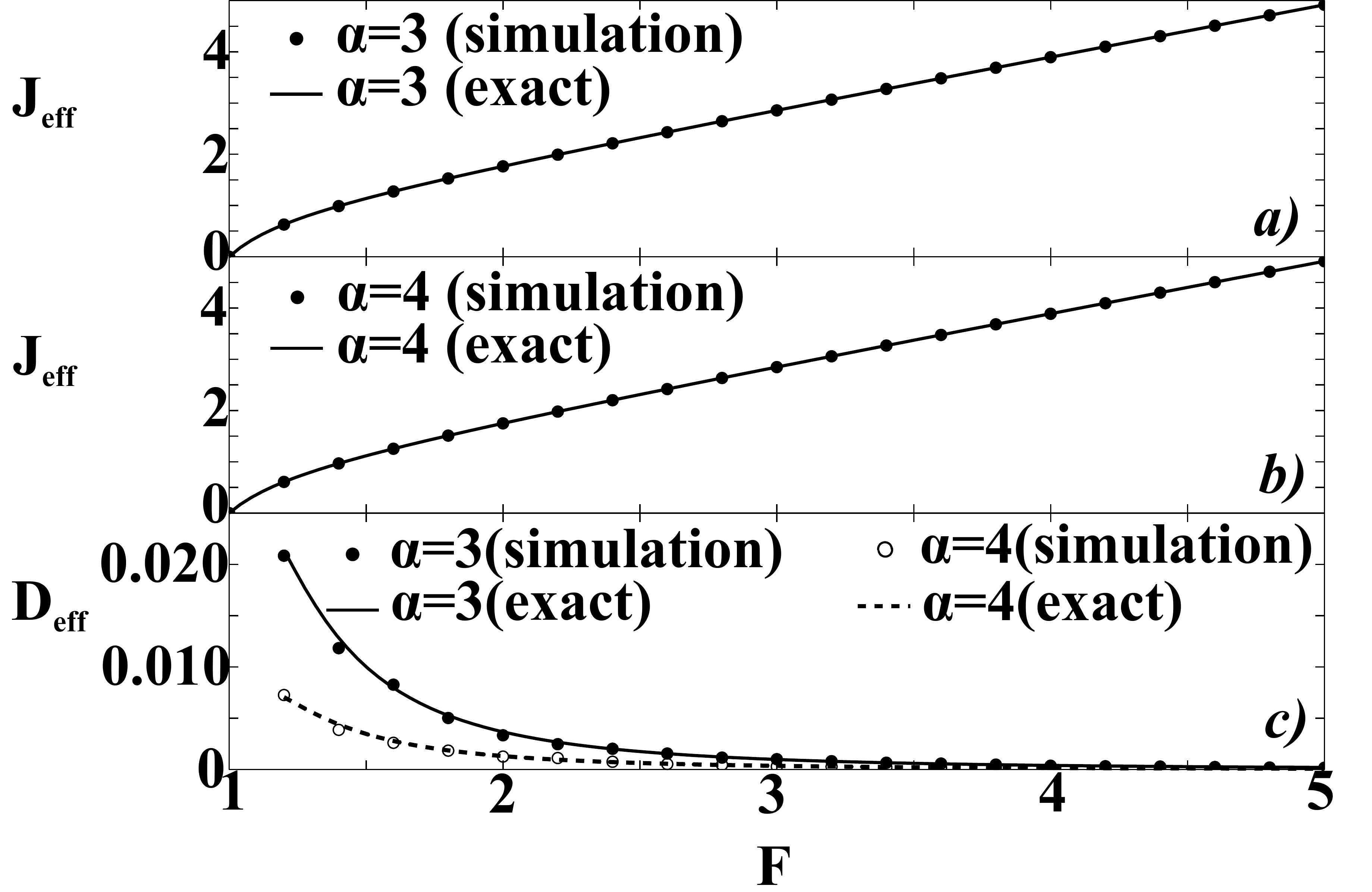}}
\end{center}
     \caption{
    Particle current and effective diffusion coefficient versus $F$.  (a) The particle current  versus the driving force for $\alpha = 3$. Solid line: analytical prediction given in Eq.~\eqref{J_Diff_Norm}. Filled circles: numerical simulation of 500 particle during a time of $10^5$ arb. units.   (b) The same as (a) but using $\alpha = 4$. (c) Effective diffusion coefficient versus $F$. We show the analytical prediction for $D_\mathrm{eff}$ given in Eq.~\eqref{D_Diff_Norm} for $\alpha =3$ (solid line) and  $\alpha =4$ (dashed line). We also display the effective diffusion coefficient obtained from the numerical simulations described in (a). We used the parameter values  $\alpha = 3$ (filled circles) and $\alpha = 4$ (open circles) to compare against the theoretical counterpart, given a good agreement within the accuracy of our numerical experiments.  
                  }
\label{fig:J_D_vs_F}
\end{figure}

In Fig.~\ref{fig:J_D_vs_F}(a) and~\ref{fig:J_D_vs_F}(b) we can appreciate the behavior of the particle current $J_{\mathrm{eff}}$ versus the driving force $F$. We plot the theoretical prediction given in Eq.~\eqref{J_Diff_Norm} for two different values of $\alpha$: for $\alpha = 3$  and $\alpha = 4$ (solid lines). We also plot the diffusion coefficient obtained by means of numerical simulations for the same cases: for $\alpha = 3$  and $\alpha = 4$ (filled circles). To estimate the particle current (and the effective diffusion coefficient) we simulated 500 particles under the dynamics defined in Eq.~\eqref{eq:motion}, each particle placed on a different realization of the random potential. The total simulation time was $10^4$ arb.~units. Then, we obtained the particle current and the diffusion coefficient estimating the mean position of the particles and its variance by averaging over the 500 trajectories obtained from the simulations. In Fig.~\ref{fig:J_D_vs_F}(c) we show the effective diffusion coefficient $D_{\mathrm{eff}}$ as a function of the driving force $F$. We plot the theoretical prediction for $D_{\mathrm{eff}}$ given in Eq.~\eqref{D_Diff_Norm} for $\alpha = 3$ (solid line) and $\alpha =4$ (dashed line). We also display the diffusion coefficient obtained from the numerical simulations described above for the estimation of the particle current. In all cases we observe good agreement with the theoretical predictions within the accuracy of our numerical simulations.

%%
%%==================== FIGURE =========================
%%
%%==================== FIGURE =========================
%

\begin{figure}[ht]
\begin{center}
\scalebox{0.25}{\includegraphics{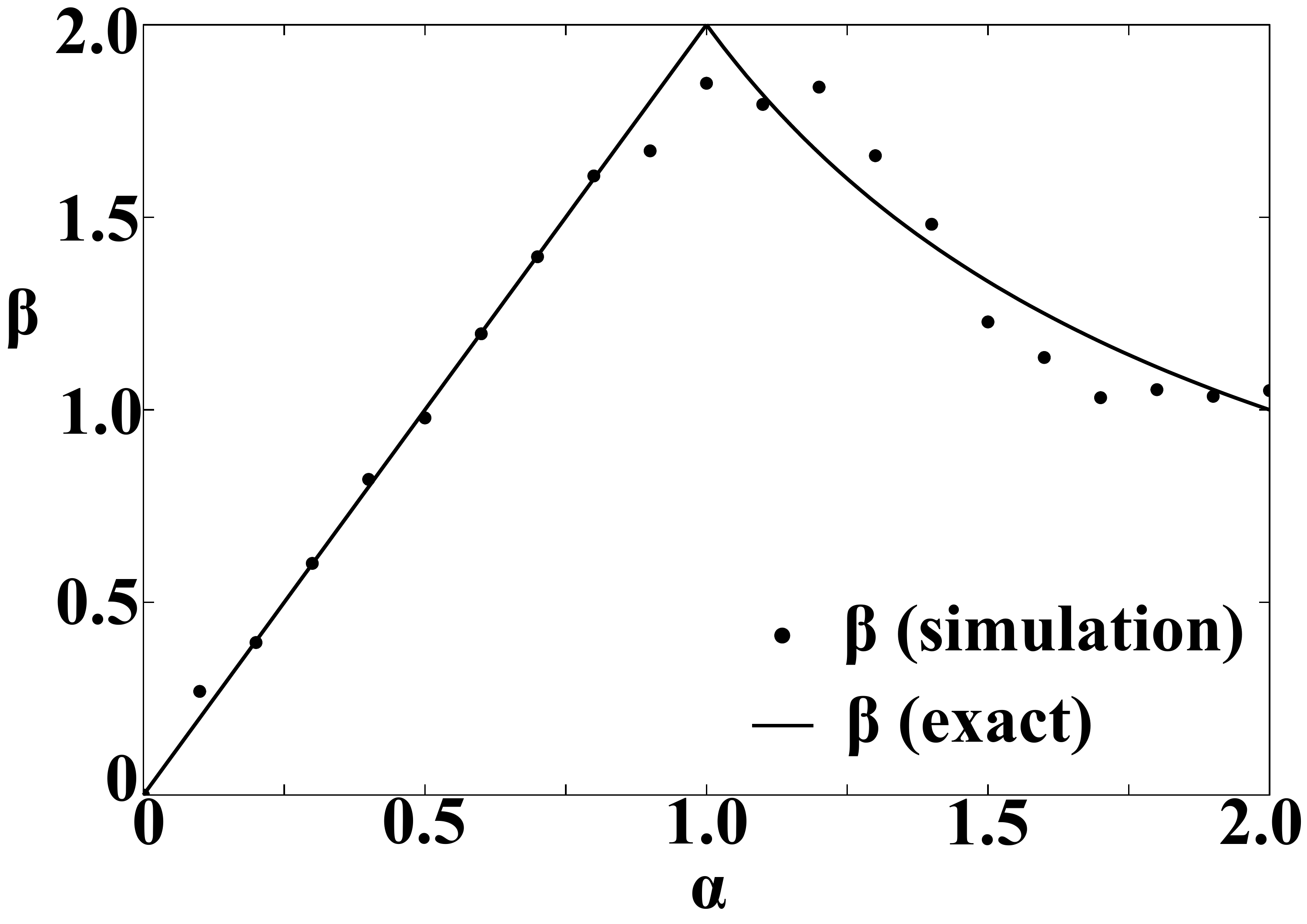}}
\end{center}
     \caption{
     Diffusion exponent versus $\alpha$. We plot the theoretical exponent $\beta$ (solid line) versus $\alpha$ obtained from Eq.~\eqref{eq:MSD} for $F=1.5$. We also plot the diffusion exponent obtained from numerical simulations (filled circles). We observe that the numerically estimated diffusion exponent is in good agreement with the theoretical result in the range $0<\alpha <1$. However, in the interval $1<\alpha <2$ the diffusion exponent exhibits large fluctuations. The last observation is due to the fact that the random variable to which converge $\Delta  X_t^2$ (for $t\to \infty$) does not have a finite mean. This particularly implies that the estimation of the mean value of $\Delta  X_t^2$ does not converge. This phenomenon is not present in $0<\alpha <1$ because the random variable $\Delta  X_t^2$ has a finite mean value.
              }
\label{fig:Anomalous_diff}
\end{figure}

In Fig.~\ref{fig:Anomalous_diff} we observe the behavior of the diffusion exponent $\beta$ versus $\alpha$. We plot the theoretical prediction for $\beta$ (solid line), given through Eq.~\ref{eq:MSD}, versus $\alpha$ for $F=1.5$. We also show the diffusion exponent obtained from numerical simulations (open circles) for the same parameter values as the theoretical curve. It is interesting to note that in the range $0<\alpha<1$ we observe that the diffusion exponent fits better to the theoretical curve than in the range $1<\alpha<2$. This phenomenon is actually a manifestation of the nature of the random variable resulting from the limit theorems. As we see from Eq.~\eqref{eq:SD_12} and~\eqref{eq:SD_01} the square displacement $\Delta X_t^2 := \left( X_t - \mathbb{E}[X_t] \right)^2$ behaves as
\begin{eqnarray}
\label{eq_SD_01}
\Delta X_t^2 &\approx& \bigg(\frac{ t^\alpha}{  \left(\frac{\gamma}{F}  \right)^\alpha  }\, \bigg)^2 (  W^{-\alpha} - \mathbb{E}[W^{-\alpha}] )^2, \  \mbox{for} \  0<\alpha<1.
\nonumber
\\
\Delta X_t^2 &\approx& 
\bigg(  \frac{\frac{\gamma}{F}  ( L+\bar{ \delta}) }{\tau_{\mathrm{c}}^{1+1/\alpha}} + \frac{1}{\tau_{\mathrm{c}}^{1/\alpha} }  \bigg)^2 t^{2/\alpha} W^2, \  \mbox{for} \  1<\alpha<2,
\nonumber
\end{eqnarray}
%%
%%==================== FIGURE =========================
%%
%%==================== FIGURE =========================
%
The main difference between these expressions for $\Delta X_t^2$ is that the corresponding mean value $\mathbb{E}[\Delta X_t^2]$ is finite for  $0<\alpha<1$, but it does not exists in the range  $1<\alpha<2$. Indeed, we have that the square displacement $\Delta X_t^2$ seen as a random variable for fixed $t$ has a distribution having a heavy tail for $1<\alpha <2$ whose mean value does not exists. Actually, the properties of the distribution of the square displacement are given by the random variable $W^2$, where $W$ has a $\alpha$-stable distribution.  These facts imply that every realization of $\Delta X_t^2$ have large fluctuations impeding the convergence of the estimator of the mean value $\mathbb{E}[\Delta X_t^2]$.  This phenomenon is of course absent in the range $0<\alpha<1$ since the random variable $(  W^{-\alpha} - \mathbb{E}[W^{-\alpha}] )^2$ has a distribution with a well-defined mean value.

%#############################################################
%#############################################################
%
\section{Conclusions}
\label{sec:conclusions}
%
%#############################################################
%#############################################################

We have introduced a simple model for deterministic diffusion which exhibit a transition from normal to anomalous diffusion. The model consists of an ensemble of non-interacting overdamped particles on a random potential under the influence of a constant driving force. The random potential can be seen as a one-dimensional medium with scarce defects which are responsible of ``delaying'' the particles.   We have show that this model is able to exhibit anomalous diffusion if the distance between defects has a heavy tailed distribution with the distribution exponent $\alpha <2$. The system also exhibits normal diffusion when the distribution exponent $\alpha \geq 2$. In the anomalous diffusive phase we observed both superdiffusion (for $1/2 < \alpha < 2$) and subdiffusion  (for $0<\alpha <1/2$). Moreover, we proved that the transport is normal (which means that the particle current is well-defined) for all the values of $\alpha$. However, we observed that the particle current versus $\alpha$ exhibits a second-order-like ``phase transition''. Explicitly, we showed that the particle current is continuous and piecewise smooth: it is a strictly decreasing function of $\alpha$ for $\alpha \geq 1$ and a constant function in the range $0< \alpha <1$. Particularly the fact that the particle current remains constant in the interval $0< \alpha <1$ seems to be a counterintuitive phenomenon. This is because we intuitively expect that the less defects in the medium the lower particle current we have. This is not the case for  $\alpha < 1$ because the particle current remains constant independently of the ``quantity'' of defects (or, equivalently, the value of $\alpha$). Finally, another phenomenon that it is worth mentioning is the fact that the square displacement $\Delta X_t^2$, seen as a random variable, has a heavy tailed distribution in the range $1< \alpha < 2$. Such a distribution is such that the mean value $\mathbb{E}[\Delta X_t^2]$ does not exists. This implies that the average of realizations (through numerical experiments) of $\Delta X_t^2$ does not converge. On the contrary, in the interval  $ 0 < \alpha < 1$ the mean value $\mathbb{E}[\Delta X_t^2]$ is finite, and by this reason, the numerical simulations fits better to the theoretical prediction for diffusion exponent. All these properties are simply consequences of the limit theorems for sums of random variables.

%#############################################################
%#############################################################
%
\section*{Acknowledgements} 
%
%#############################################################
%#############################################################

The authors thank CONACyT for financial support through Grant No. CB-2012-01-183358.

\nocite{*}

\bibliography{ScarceDefects}% Produces the bibliography via BibTeX.

\end{document}